\documentclass[aps,prl,twocolumn,preprintnumbers,10pt,showpacs]{revtex4-1}
\usepackage{amsmath,bm,amssymb,color}

\usepackage{graphicx}

\newcommand{\dsigma}{{\rm d}\sigma}
\def\gapprox{\lower .7ex\hbox{$\;\stackrel{\textstyle >}{\sim}\;$}}
\newcommand\roots{8}
\newcommand\jet{{\rm~jet}}

\allowdisplaybreaks

\begin{document}

\preprint{IPPP/15/44, ZU-TH 23/15}

\title{Precise QCD predictions for the production of a $Z$ boson in association with a hadronic jet}

\author{A.\ Gehrmann-De Ridder$^{a,b}$, T.\ Gehrmann$^a$, E.W.N.\ Glover$^c$, A.\ Huss$^{a,b}$, T.A.\ Morgan$^c$}

\affiliation{
$^a$Department of Physics, University of Z\"urich, CH-8057 Z\"urich, Switzerland\\
$^b$Institute for Theoretical Physics, ETH, CH-8093 Z\"urich, Switzerland\\
$^c$Institute for Particle Physics Phenomenology, Department of Physics, University of Durham, Durham, DH1 3LE, UK}

\pacs{12.38Bx}

\begin{abstract}
We compute the cross section and differential distributions for the production of a $Z$ boson in  association with a hadronic jet to next-to-next-to-leading order (NNLO) in perturbative QCD, including the leptonic decay of the $Z$ boson. We present numerical results for the transverse momentum and rapidity distributions of both the $Z$ boson and the associated jet at the LHC. We find that the NNLO corrections increase the next-to-leading order (NLO) predictions by approximately 1\% and significantly reduce the scale variation uncertainty.
\vspace*{1.5em}
\end{abstract}

\maketitle


The Drell--Yan production of lepton pairs is a benchmark process at hadron colliders like the Large Hadron Collider (LHC). The production of  $Z$ bosons (or off-shell photons) with subsequent leptonic decays has both a clean and readily identifiable signature and a large event rate. It is a key process for precision measurements of electroweak (EW) parameters, and also 
allows to probe various aspects of the strong interaction, including parton 
 distribution functions (PDFs), the strong coupling constant $\alpha_s$, and the behaviour of processes involving multiple 
 scales. 
It is moreover a key ingredient in calibrating several parts of the detector (including the jet energy scale) and can potentially be used to measure the luminosity of the collider. At the LHC, the  $Z$ boson is almost always produced together with additional QCD radiation thereby providing a perfect testing ground for our theoretical understanding of both strong and electroweak physics in a hadronic environment. Together the combination of precise experimental data and reliable theoretical predictions enables a variety of precision measurements at the LHC~\cite{ATLASZJ,CMSZJ}. 


The importance of the neutral current Drell--Yan process is also reflected in the effort to make the theoretical predictions as precise as possible. For inclusive $Z$ production, theoretical predictions at per-cent level accuracy are available. To attain this 
level of precision, a variety of 
higher-order corrections in QCD and the EW theory had to be considered. 
The cross section for $Z$ production is known at next-to-next-to-leading order accuracy (i.e.\ at two loops) with respect to QCD corrections~\cite{dyNNLO}. Corrections beyond this order have been studied in the soft-virtual 
approximation~\cite{dyN3LO}.  The NNLO QCD corrections have been combined with a resummation of next-to-next-to-leading logarithmic effects~\cite{dyresum} which is necessary to predict the transverse momentum distribution of the $Z$ boson at small $p_T$ and matched with parton showers~\cite{dyNNLOPS}. In the electroweak theory, 
the next-to-leading order 
corrections~\cite{dyNLOEW,dyNLOEWPS} and the mixed QCD--EW corrections~\cite{dyQCDEW} also contribute to the 
precise description of this process.  
 Drell--Yan production in association with hadronic jets has also been intensively studied. The NLO QCD corrections for $Z$ + 1 jet~\cite{ZJNLO}, $Z$ + 2 jets~\cite{ZJJNLO}, $Z$ + 3 jets~\cite{Z3JNLO} and  $Z$ + 4 jets~\cite{Z4JNLO} are known while the NLO EW corrections for $Z$ + 1 jet~\cite{ZJNLOEW} and $Z$ + 2 jets~\cite{Z2JNLOEW} have also been derived.   

In this letter, we report on the calculation of the NNLO contributions to the neutral-current Drell--Yan process in which the dilepton pair is produced in association with a hard, visible hadronic jet, 
$$
pp \to Z/\gamma^* + \jet \to \ell^+\ell^- + \jet + X.
$$
Our results are obtained in the form of a parton-level event generator that provides the corrections in 
a fully differential form, including the $Z/\gamma^*$ boson decay to two charged leptons. The final state of the 
hard-scattering process is completely reconstructable and the application of an invariant mass cut on the lepton pair can 
ensure that the process is dominated by resonant $Z$ bosons.

\begin{figure}[t]
(a)\includegraphics[width=2.2cm]{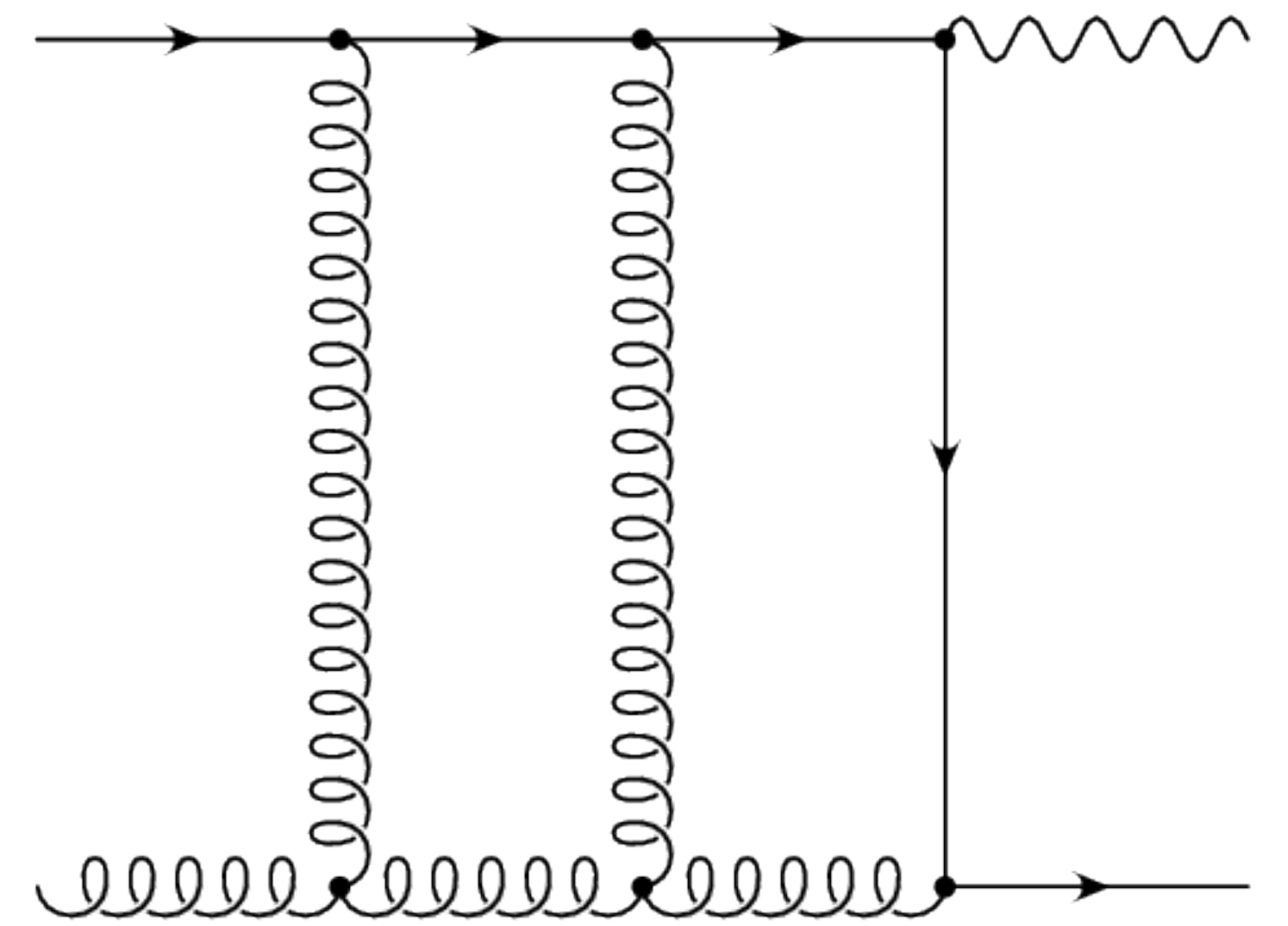}
(b)\includegraphics[width=2.2cm]{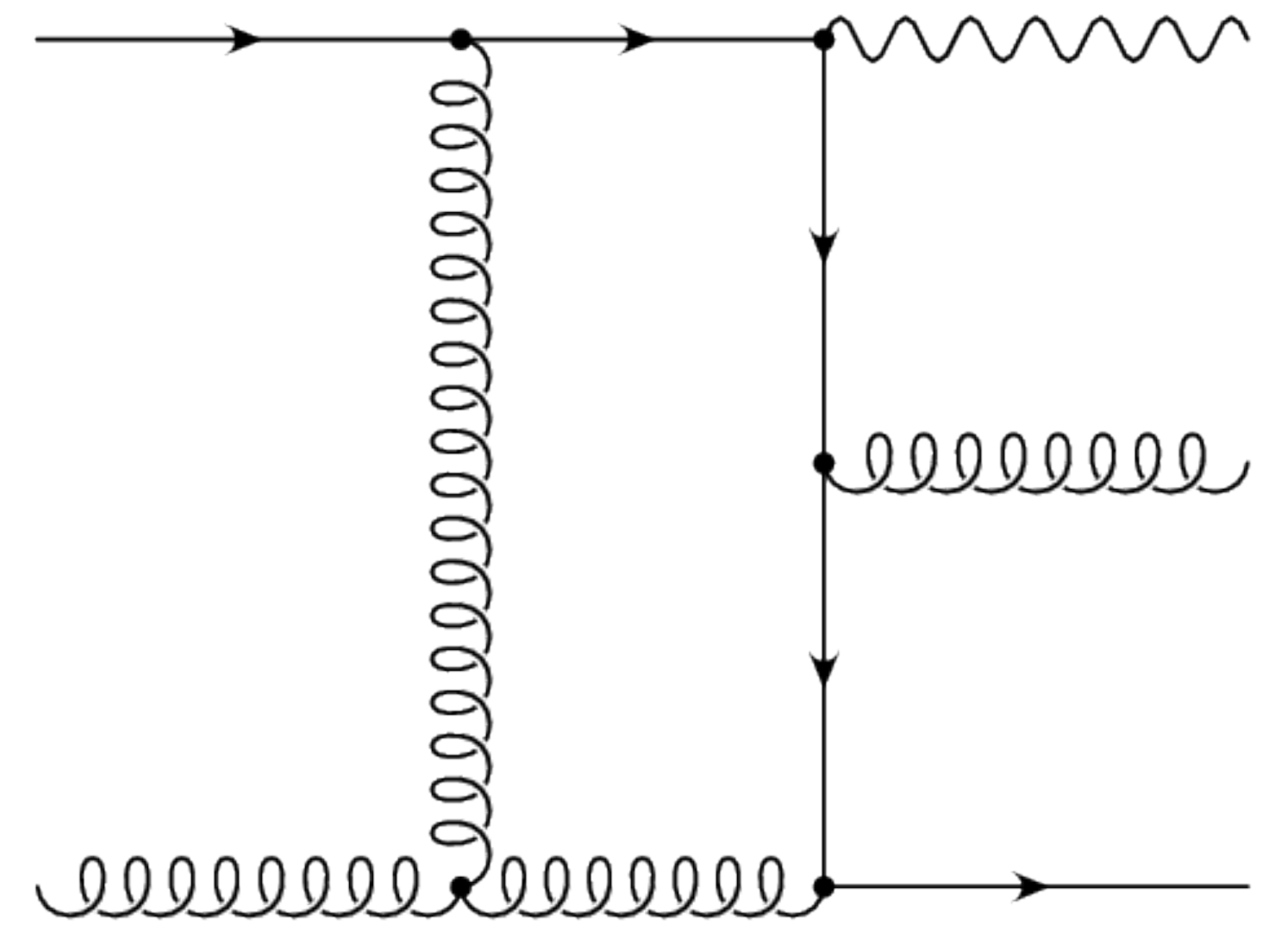}
(c)\includegraphics[width=2.2cm]{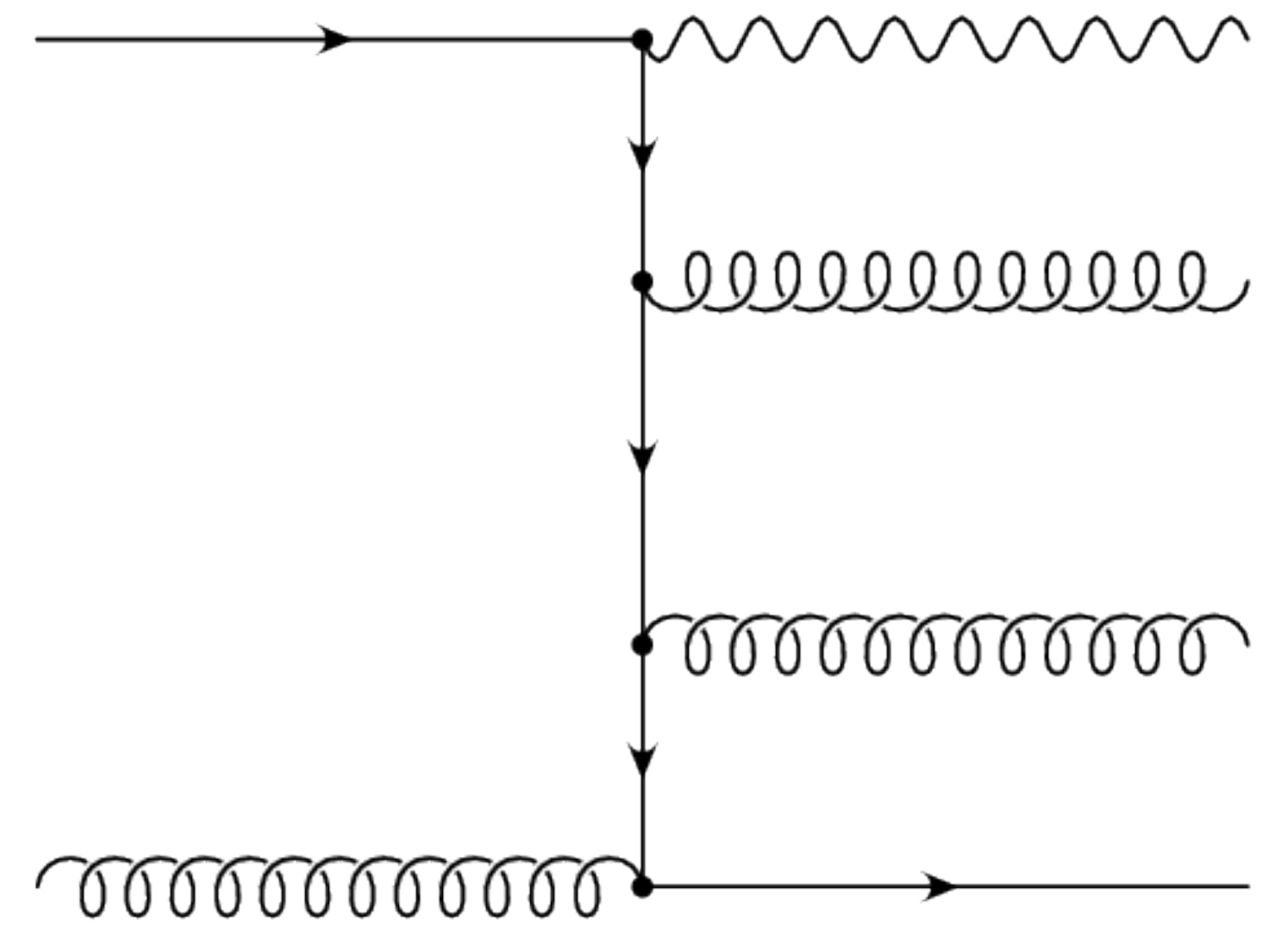}
\caption{Representative Feynman diagrams for  
(a) two-loop $Z$ boson-plus-three-parton amplitudes (b)
one-loop $Z$ boson-plus-four-parton amplitudes and (c) 
tree-level $Z$ boson-plus-five-parton amplitudes.\label{fig:FD}}
\end{figure}

The NNLO corrections to $Z$ boson + jet production  in hadronic collisions receive contributions from three types of parton-level processes:  
(a) the two-loop corrections to $Z$ boson-plus-three-parton processes~\cite{Z3p2l}, 
(b) the one-loop  corrections to $Z$ boson-plus-four-parton processes~\cite{Z4p1l,ZJJNLO} and 
(c)  the tree-level $Z$ boson-plus-five-parton  processes~\cite{Z5p0l,ZJJNLO}. 
Figure~\ref{fig:FD} shows representative Feynman diagrams for each of the partonic multiplicities. The ultraviolet renormalized matrix elements for these processes are integrated  over the final state phase space appropriate to $Z$ boson + jet final states. All three types of contributions  are infrared-divergent and only their sum is finite.  While infrared divergences from the virtual corrections are explicit in the  one-loop and two-loop matrix elements, divergences from unresolved real radiation become explicit only after phase space integration. The divergences are regulated using dimensional regularization, and a variety of methods have been used for their extraction from the real radiation  contributions. All these methods are based on the isolation of the divergent configurations, which are then integrated over the phase space and added to the virtual corrections to yield a finite result:  
sector decomposition~\cite{secdec}, sector-improved residue subtraction~\cite{stripper}, antenna subtraction~\cite{ourant}, $q_T$-subtraction~\cite{qtsub} and  N-jettiness subtraction~\cite{njettiness} have all been applied successfully in the calculation of NNLO corrections for a range of LHC processes.  

In this calculation we employ the antenna subtraction method~\cite{ourant} in which the  real radiation subtraction terms are constructed from antenna functions.  These antenna functions capture all the unresolved
radiation emitted between a pair of hard
radiator partons. For hadron-collider observables, either hard radiator can be in the  initial or final state, and all unintegrated and integrated  antenna functions were derived in Refs.~\cite{hadant,gionata,monni,ritzmann}.  The cross section corresponding to an initial partonic state $ij$ is given by, 
\begin{eqnarray}
\dsigma_{ij,NNLO}&=&\int_{{\rm{d}}\Phi_{3}}\left[\dsigma_{ij,NNLO}^{RR}-\dsigma_{ij,NNLO}^S\right]
\nonumber \\
&+& \int_{{\rm{d}}\Phi_{2}}
\left[
\dsigma_{ij,NNLO}^{RV}-\dsigma_{ij,NNLO}^{T}
\right] \nonumber \\
&+&\int_{{\rm{d}}\Phi_{1}}\left[
\dsigma_{ij,NNLO}^{VV}-\dsigma_{ij,NNLO}^{U}\right],
\end{eqnarray}
where each of the square brackets is finite and well 
behaved in the infrared singular regions. 
The construction of the subtraction terms $\dsigma_{ij,NNLO}^{S,T,U}$ follows 
closely the procedure established for jet production~\cite{joao} and Higgs~+~jet production~\cite{ourHJ}. Powerful checks of our formalism are that (a) the poles in the dimensional regularization parameter $\epsilon$ cancel analytically and (b) that the subtraction terms accurately reproduce the singularity structure of the real radiation matrix elements.
 
Using the antenna subtraction method, we have derived the corresponding subtraction terms for all partonic initial states and all color contributions for $Z$ boson-plus-jet production through to NNLO and implemented them in a parton-level event generator. With this program, we can  compute any infrared safe observable related to $Z$ + jet final states to NNLO accuracy. The $Z$ boson decay to two charged leptons is included, such that realistic event selection cuts on the leptonic final state can be  applied. Renormalization and factorization scales can be chosen (dynamically) on an event-by-event basis. 

For our numerical computations, we use the NNPDF2.3 parton distribution functions~\cite{nnpdf}  
with the corresponding value of $\alpha_s(M_Z)=0.118$ at NNLO, and $M_Z=91.1876~$GeV. Note that we systematically use the 
same set of PDFs and the same value of $\alpha_s(M_Z)$ for the LO, NLO and NNLO predictions.  The factorization and renormalization scales are chosen to be $\mu \equiv \mu_F=\mu_R=M_Z$, 
with a theoretical uncertainty estimated by varying the scale choice by a factor in the range $[1/2,2]$. 

We require that the leptons have pseudorapidity, $|\eta^{\ell}| < 5$ and that the dilepton invariant mass is close to the $Z$ boson mass, $80$~GeV $< m_{\ell\ell} < 100$~GeV. Jets are reconstructed  using the anti-$k_T$ algorithm~\cite{antiKT} with $R=0.5$ and are required to have $p_T^{\rm jet} >30$~GeV and $|y^{\rm jet}| < 3$.
With these cuts, we find that the total cross section at different perturbative orders is given by,
\begin{eqnarray}
\sigma_{LO} &=& 103.6^{+7.7}_{-7.5}~\mbox{ pb}\;,\nonumber \\
\sigma_{NLO} &=& 144.4^{+9.0}_{-7.2}~\mbox{ pb}\;,\nonumber \\
\sigma_{NNLO} &=& 145.8^{+0.0}_{-1.2}~\mbox{ pb}\;
\end{eqnarray}
so that the inclusive NNLO corrections amount to a 1\% increase on the NLO cross section.

More information on the impact of the NNLO QCD corrections can be gained from differential distributions in the 
kinematical variables of the $Z$ boson and the jet. 
In the kinematical distributions and ratio plots, the error band describes the scale variation envelope as described above, where the denominator in the ratio plots is evaluated at fixed central
scale, such that the band only reflects the variation of the numerator.
\begin{figure}[t]
\centering
\includegraphics[angle=0,width=\linewidth]{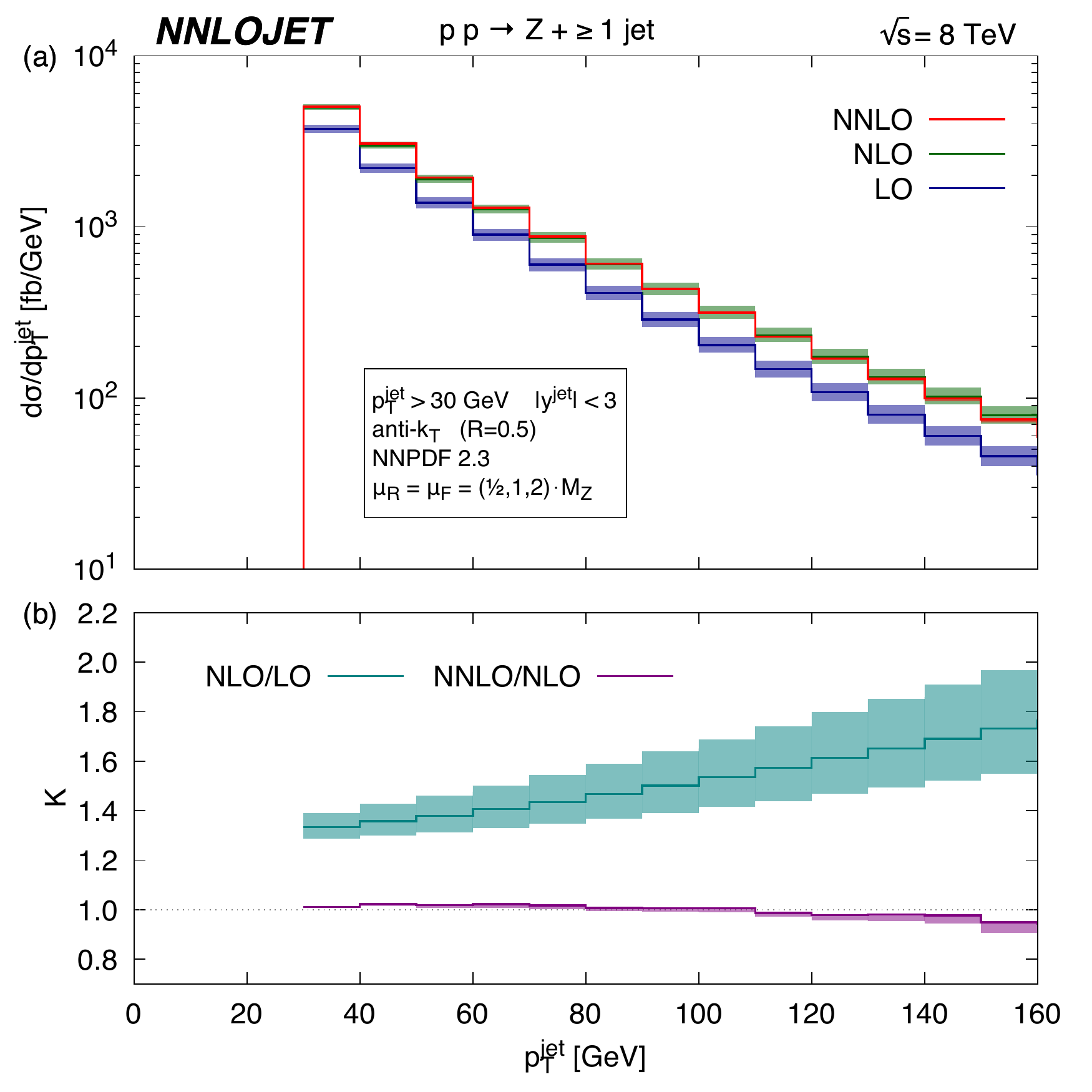}
\caption{(a) Transverse momentum distribution of the leading jet
in inclusive $Z+1\jet$ production  in $pp$ collisions with 
$\sqrt{s}=\roots$~TeV at LO (blue), NLO (green), NNLO (red) 
and (b) Ratios of different perturbative orders, NLO/LO (turquoise) and NNLO/NLO (mauve).
\label{fig:ptj1}}
\end{figure}
Figure~\ref{fig:ptj1}(a) shows the inclusive leading jet transverse energy distribution in 10~GeV bins. Due to the inclusiveness 
of the observable, events with two or three jets 
with $p_T^{\rm jet} >30$~GeV and $|y^{\rm jet}| < 3$
are also included. The relative corrections are further exposed in Figure~\ref{fig:ptj1}(b) where we show the ratio,
$K = {\dsigma^{(N)NLO}(\mu)}/{\dsigma^{(N)LO}(\mu=M_Z)}$. The band shows the effect of varying $\mu$ in the range $[1/2,2] M_Z$ in the numerator while keeping $\mu=M_Z$ in the denominator. For our set of cuts and input parameters, we see that the NLO corrections increase the cross section by between 30\% to 70\%. At low transverse momentum the NNLO corrections are a positive correction of approximately 1\%. The variation with the unphysical scales is significantly reduced as we move from NLO to NNLO.  

\begin{figure}[t]
\centering
\includegraphics[angle=0,width=\linewidth]{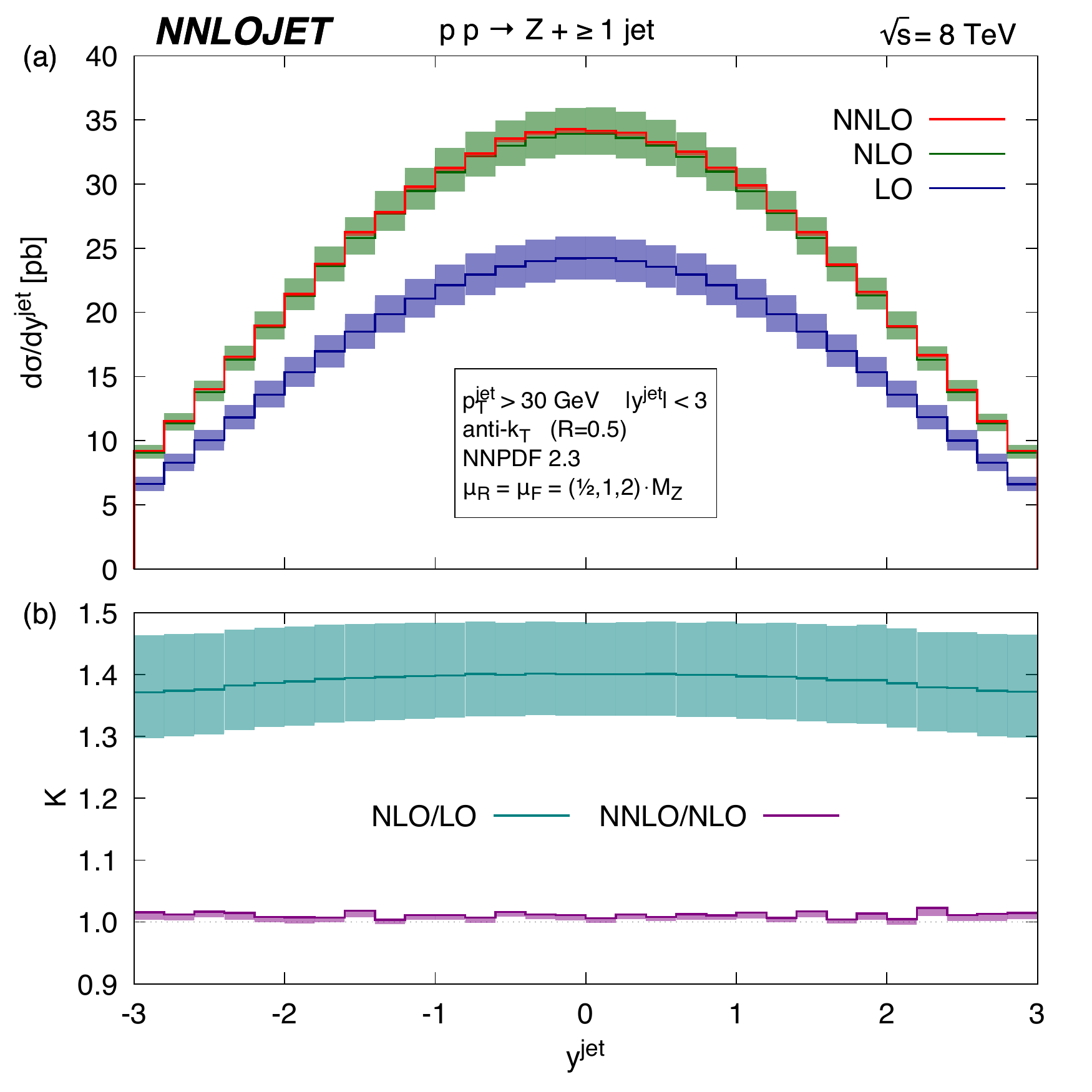}
\caption{
(a) Rapidity distribution of the leading jet in inclusive $Z+1\jet$ production
 in $pp$ collisions with 
$\sqrt{s}=\roots$~TeV at LO (blue), NLO (green), NNLO (red) 
and (b) Ratios of different perturbative orders, NLO/LO (turquoise) and NNLO/NLO (mauve).
\label{fig:etaJ}}
\end{figure}
The rapidity distribution of the 
leading jet is displayed in Figure~\ref{fig:etaJ}.
Note that the distribution is restricted by the requirement that $|y^{\rm jet}| < 3$.
We see that the NLO corrections are typically 35\%--40\% and
relatively flat.  The NNLO corrections increase the cross section by approximately 1\% over the whole range of $y^{\rm jet}$ with a significantly reduced scale dependence.

\begin{figure}[t]
\centering
\includegraphics[angle=0,width=\linewidth]{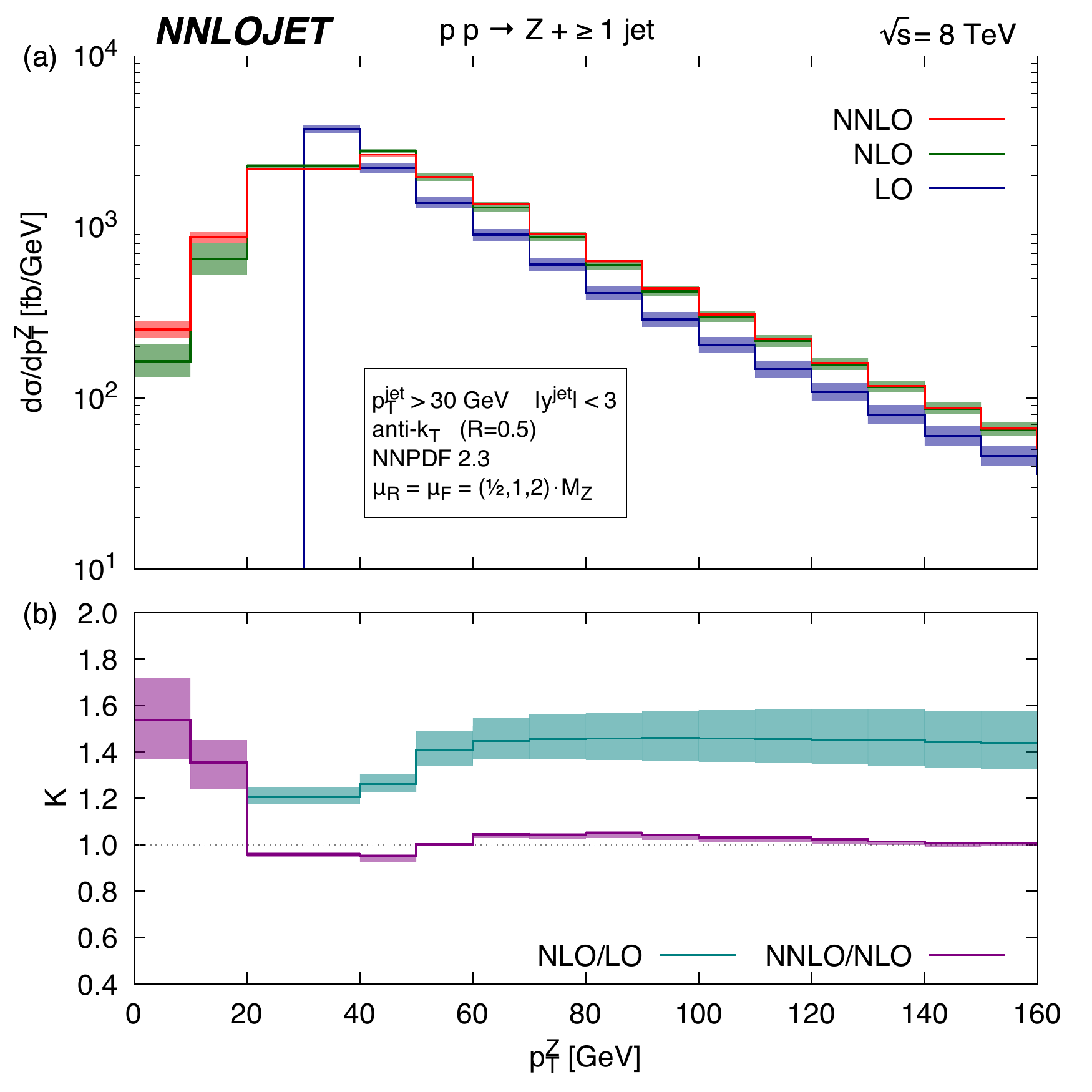}
\caption{(a) Transverse momentum distribution of the $Z$ boson
in inclusive $Z+1\jet$ production in $pp$ collisions with 
$\sqrt{s}=\roots$~TeV at LO (blue), NLO (green), NNLO (red) 
and (b) Ratios of different perturbative orders, NLO/LO (turquoise) and NNLO/NLO (mauve).
\label{fig:ptV}}
\end{figure}
The $Z$ boson $p_T$ distribution in inclusive $Z$ + jet production 
is shown in Figure~\ref{fig:ptV} where we observe an 
interesting structure around $p_T^Z \sim 30$~GeV. 
This behaviour arises from the fact that the $Z$ boson is recoiling against 
a complicated hadronic final state that contains at least one jet with $p_T^{\rm jet} > 30$~GeV. 
For this set of cuts, the leading order process is constrained to have $p_T^Z > 30$~GeV, while higher order real radiation 
corrections lift this limitation, since extra partonic radiation can also balance the transverse momentum of the leading jet. 
This Sudakov shoulder phenomenon is also observed in H+jet production~\cite{ourHJ,HJall}; it is well understood~\cite{sudakov} and leads to large higher order corrections, which require logarithmic resummation.
Nevertheless, the NNLO corrections tend to stabilise the NLO result, and in fact simply represent a 
NLO correction to the $p_T^{Z}$ distribution for $Z$ + jet events in this region. At larger transverse momenta, the 
NNLO corrections increase the prediction by approximately 1\%. 
 
\begin{figure}[t]
\centering
\includegraphics[angle=0,width=\linewidth]{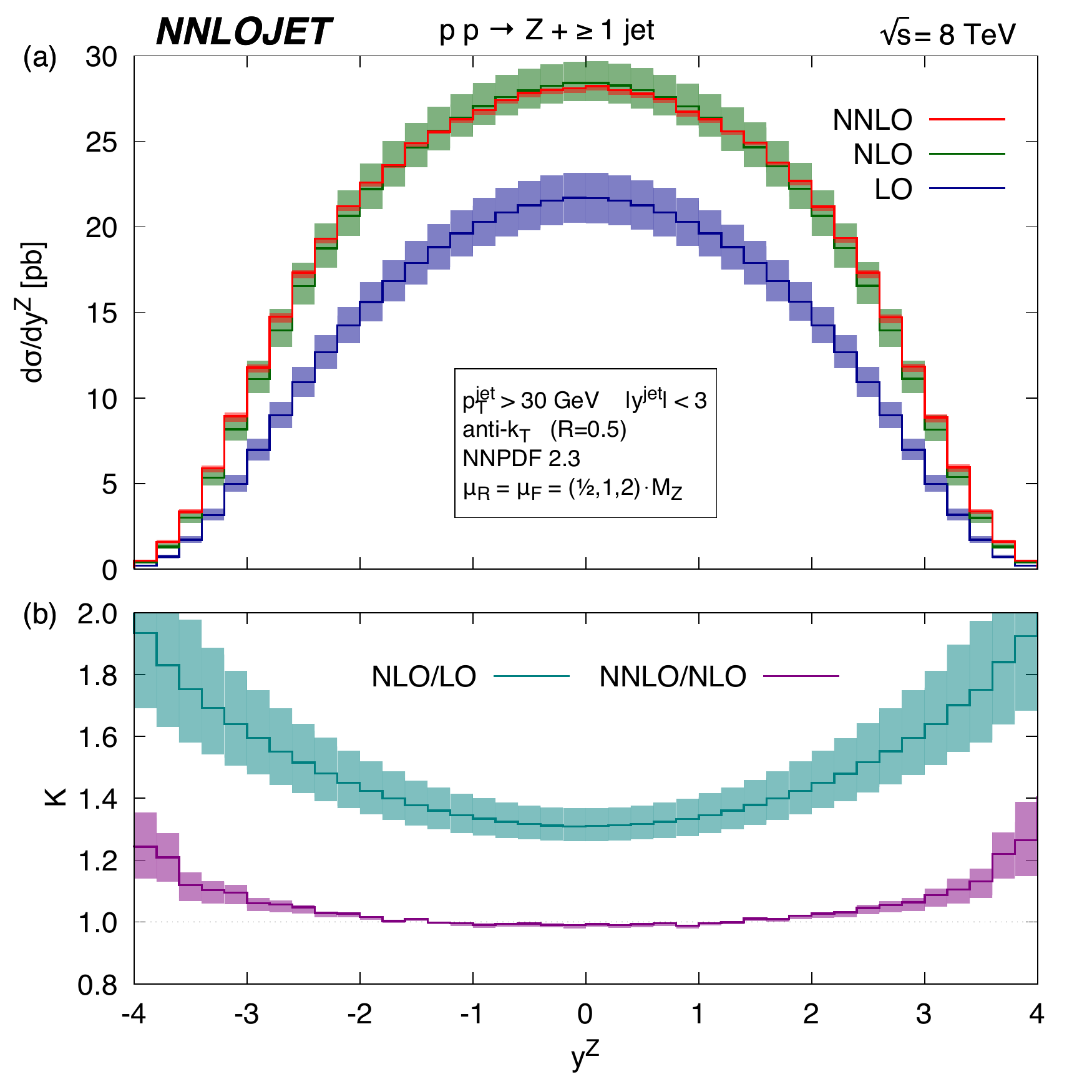}
\caption{(a) Rapidity distribution of the Z boson in inclusive $Z+1\jet$ production 
 in $pp$ collisions with 
$\sqrt{s}=\roots$~TeV at LO (blue), NLO (green), NNLO (red) 
and (b) Ratios of different perturbative orders, NLO/LO (turquoise) and NNLO/NLO (mauve).
\label{fig:etaV}}
\end{figure}
Figure~\ref{fig:etaV} shows the rapidity distribution of the $Z$ boson. The NLO and NNLO corrections are largest in the forward/backward regions where the phase space is enlarged by the possibility that the hadronic radiation partially balances leading to a smaller $Z$ $p_T$.  In these regions, one of the parton momentum fractions is reaching a maximal value.  In the central region, the NNLO corrections are very small with a reduced scale dependence.

In the differential distributions  we observe that the corrections are not always
uniform, implying that a rescaling of lower-order predictions is insufficient for precision applications. 
The need for using the fully differential higher order predictions can be understood for example  
 in the extraction of parton distributions functions from $Z$ + jet production. 
At leading order, the momentum fractions of the incoming partons is completely fixed by the transverse momenta and rapidities of the final state particles.
At higher orders, the real radiation spoils the leading order kinematics,  such that
\begin{eqnarray*}
x_1 &\ge& \frac{1}{\sqrt{s}} \left (\sqrt{(p_T^Z)^2+m^2_{\ell\ell}}\,\exp(\phantom{-}y_Z) + p_T^{\rm jet}\,\exp(\phantom{-}y_{\rm jet}) \right),\\
x_2 &\ge& \frac{1}{\sqrt{s}} \left (\sqrt{(p_T^Z)^2+m^2_{\ell\ell}}\,\exp(-y_Z) + p_T^{\rm jet}\,\exp(-y_{\rm jet})  \right),
\end{eqnarray*}
where the equality is restored only for the leading order kinematics ($p_T^Z = p_T^{\jet}$).
The relevant $x$ ranges probed by $Z$ boson-plus-jet production is thus determined by the transverse momentum and rapidity distribution of the $Z$ boson and the jet. For our cuts, the smallest momentum fractions probed are $x \sim 8\cdot 10^{-3}$, and 
smaller values of $x$ can be attained by enlarging the rapidity interval or by lowering the transverse momentum cut. 

In this manuscript we have presented the complete NNLO QCD calculation of $Z$ boson production in association with a jet in hadronic collisions including all partonic subprocesses. This process is measured experimentally to high precision~\cite{ATLASZJ,CMSZJ} and is an important ingredient to a variety of 
precision studies of Standard Model parameters and derived quantities as well as a key element in the LHC detector 
calibration. 
We have achieved this using the antenna subtraction method that has been successfully applied to other processes at the LHC. For all of the observables considered here, we observed a very significant reduction of the respective uncertainties in the theory prediction due to variations of the factorization and renormalization scales with a residual NNLO scale uncertainty of around 1\% on the normalization of the distributions.   
Our calculation will be a crucial tool for precision studies of $Z$ boson + jet production in the upcoming data taking periods at the CERN LHC.

This research was supported in part by the Swiss National Science Foundation (SNF) under contracts 200020-149517 
and CRSII2-141847, in part by
the UK Science and Technology Facilities Council as well as by the Research Executive Agency (REA) of the European Union under the Grant Agreement PITN-GA-2012-316704  (``HiggsTools''), and the ERC Advanced Grant MC@NNLO (340983).

{\bf Note added:} After this paper was initially submitted to Physical Review Letters, a second calculation (employing a different subtraction scheme) of Z+jet production at NNLO precision has been presented and published~\cite{Boughezal:2015ded}. In coordination with the authors of~\cite{Boughezal:2015ded}, we performed an in-depth comparison, by running our code with their settings (cuts, parton distributions, scale choice). This comparison uncovered an error in the numerical code used in~\cite{Boughezal:2015ded}, which alters their published results. After correction of this error, the code developed in~\cite{Boughezal:2015ded} agrees with our results.

\end{document}